\documentclass[12pt,a4paper]{iopart}

\usepackage{rotating,graphicx}
\usepackage{amsfonts}%
\usepackage{caption}
\usepackage[utf8]{inputenc}
\usepackage[english]{babel} 
\usepackage[T1]{fontenc}
\usepackage{color}
\usepackage[colorlinks]{hyperref}
\usepackage{setspace}
\usepackage{layouts}
\begin{document}

\title[Towards a quantitative description of filamentary SOL transport] {Recent progress towards a quantitative description of filamentary SOL transport}
\author{D. Carralero$^{1}$, M. Siccinio$^{1}$, M. Komm$^{2}$, S. A. Artene$^{1,3}$, F. A. D'Isa$^{1}$, J. Adamek$^{2}$, L. Aho-Mantila$^{4}$, G. Birkenmeier$^{1,3}$, M. Brix$^{5}$, G. Fuchert$^{1,6}$,  M. Groth$^{7}$, T. Lunt$^{1}$, P. Manz$^{1,3}$, J. Madsen$^{8}$, S. Marsen$^{5,6}$,  H. W. Müller$^{1,9}$ U. Stroth$^{1}$, H. J. Sun$^{1}$, N. Vianello$^{10,11}$, M. Wischmeier$^{1}$, E. Wolfrum$^{1}$, ASDEX Upgrade Team$^{1}$, COMPASS Team$^{2}$, JET Contributors$^{12}$ and the EUROfusion MST team$^{13}$.}
\address{$^1$ Max-Planck-Institut für Plasmaphysik, EURATOM Association, 85748 Garching, Germany.Austria}
\address{$^2$ Institute of Plasma Physics AS CR, Prague, Czech Republic.}
\address{$^3$ Physik-Department E28, Technische Universität München, Garching, Germany.}
\address{ $^4$ VTT Technical Research Center of Finland, Helsinki, Finland.}
\address{ $^5$ EUROfusion Consortium, JET, Culham Science Centre, Abingdon, OX14 3DB, UK.}
\address{ $^6$ Max-Planck-Institut für Plasmaphysik, Greifswald, Germany. }
\address{$^7$ Aalto University, Espoo, Finland.}
\address{ $^8$ The Technical University of Denmark, Department of Physics, DK-2800 Kgs. Lyngby, Denmark.}
\address{$^9$ Institute of Materials Chemistry and Research, University of Vienna, Währingerstrasse 42, A-1090 Vienna.}
\address{ $^{10}$ Consorzio RFX, Associazione Euratom-ENEA sulla fusione, C.so Stati Uniti 4,I-35127 Padova, Italy.}
\address{ $^{11}$ Ecole Polytechnique Fédérale de Lausanne, Centre de Recherches en Physique des Plasmas, Lausanne, Switzerland. }
\address{$^{12}$ See the author list of “Overview of the JET results in support to ITER” by X. Litaudon et al. to be published in Nuclear Fusion Special issue: overview and summary reports from the 26th Fusion Energy Conference (Kyoto, Japan, 17-22 October 2016).}
\address{ $^{13}$ See http://www.euro-fusionscipub.org/mst1.}
\ead{daniel.carralero@ipp.mpg.de}


\begin{abstract}
A summary of recent results on filamentary transport, mostly obtained in the ASDEX-Upgrade tokamak (AUG), is presented and discussed in an attempt to produce a coherent picture of SOL filamentary transport: A clear correlation is found between L-mode density shoulder formation in the outer midplane and a transition between the sheath limited and the inertial filamentary regimes. Divertor collisionality is found to be the parameter triggering the transition. A clear reduction of the ion temperature takes place in the far SOL after the transition, both for the background and the filaments. This coincides with a strong variation of the ion temperature distribution, which deviates from Gaussianity and becomes dominated by a strong peak below $5$ eV. The filament transition mechanism triggered by a critical value of collisionality seems to be generally applicable to inter-ELM H-mode plasmas, although a secondary threshold related to deuterium fueling is observed. EMC3-EIRENE simulations of neutral dynamics show that an ionization front near the main chamber wall is formed after the shoulder formation. Finally, a clear increase of SOL opacity to neutrals is observed associated to the shoulder formation. A common SOL transport framework is proposed account for all these results, and their potential implications for future generation devices are discussed.
\end{abstract}

\section{Introduction}\label{intro}

Heat and particle transport onto plasma-facing components of fusion devices is determined in the Scrape-off Layer (SOL) by the balance between parallel and perpendicular transport. While parallel heat transport is mostly dominated by conduction, the most important contribution to perpendicular transport is the perpendicular advection associated to coherent structures known as filaments. This is a key issue for next generation tokamaks, as the filaments will determine the erosion levels and the heat loads at the main chamber first wall. Also, by spreading power across the field lines, these structures may potentially reduce concentrated heat loads at the divertor strike point. Basic models for filaments \cite{Krash01} describe their $E \times B$ propagation as the result of a polarization caused by curvature drifts \cite{Garcia06}. In order to keep charge conservation, conventional models had this polarization compensated by a parallel current flowing along the filament into the wall \cite{Krash07}. This is known as the Sheath Limited regime (SL). It was later proposed that some mechanisms, such as a large increase of SOL collisionality, might electrically disconnect the midplane from the targets, leading to the so-called inertial regime (IN)\cite{Garcia06}. The transition from SL to IN regime could have important global implications, as it would cause larger filaments \cite{Krash07,Myra06} and lead to an enhanced perpendicular transport \cite{Russel07}, thus potentially increasing particle and heat loads on the first wall\cite{Birkenmeier15}. A well known example of this would be the onset of the density profile flattening known in the literature as the density “shoulder”, reported in many tokamaks when a certain density threshold is exceeded during L-mode operation \cite{Labombard01,Rudakov05,Garcia07}, and which has been explained as the result of filament disconnection from the target \cite{Dippolito06}.\\

Previous work carried out on AUG \cite{Carralero14} already established the relation between shoulder formation and a clear transition in the properties of filaments. Both phenomena shared a common threshold, which coincided with the point where collisions disconnected the midplane from the divertor target. Such a disconnection was defined using the effective collisionality criterion proposed by Myra et al. \cite{Myra06b}, $\Lambda > 1$, where $\Lambda = \frac{L/c_s}{1/\nu_{ei}}\frac{m_e}{m_i}$, and $L$ is the parallel scale length, $c_s$ the sound speed and $\nu_{ei}$ the electron–ion collision frequency. Later work confirmed these results in a multimachine study including the tokamaks JET, AUG and COMPASS\cite{Panek16}, which due to their similar magnetic configuration and difference in size form a “stepladder” to ITER \cite{Carralero14b}: In the first two devices, the density was high enough to achieve $\Lambda >1$, and both the filament transition and the shoulder formation were observed. In COMPASS, where $\Lambda >1$ was not achieved, filaments and density profiles remained constant for the whole range of observed densities. Still, it could not be decided if the process was triggered by the density – as suggested by previous literature -, the collisionality in the midplane, $\Lambda_{mid}$ (defined using far SOL $T_e$ and $n_e$ values and the connection length as L), or the local collisionality at the divertor, $\Lambda_{div}$ (defined using the average $T_e$ and $n_e$ measured around $\rho = 1.025$ at the divertor target and $1/5$ times the connection length as $L$).\\

In this work, we present the experimental and numerical effort that has been carried out in recent years with the purpose of understanding the basic mechanism determining filamentary transport in order to improve current estimations of first wall loads for ITER and DEMO. Most of this work has been done in L-mode plasmas, which allow for greater diagnostic coverage. However, complementary experiments have also been carried out in which the conclusions achieved in L-mode experiments are expanded into the more relevant H-mode. In order to interpret the experimental results, this effort has been complemented by numerical simulations: Using the EMC3-EIRENE code, L-mode discharges  featuring low and high SOL collisionality have been simulated, displaying good agreement with available measurements, and providing information out of the reach of available diagnostics, such as ionization and neutral density profiles in the outer midplane. Finally, all the presented work is discussed, aiming to provide a coherent picture of the effect of filamentary transport on the SOL, and its potential implications for next generation devices. The paper is organized as follows: In section 2, the parameter triggering the transition is determined. In section 3, a study on the scaling of filaments is presented. In section 4, the impact on SOL temperatures is discussed. In section 5, previous results are extended to H-mode. In section 6, EMC3-EIRENE simulations are presented and discussed. In section 7, common SOL transport framework is proposed account for all these results, and their potential implications for future generation devices are discussed. Finally, the main conclusions are outlined in section 8.\\

\section{The role of divertor collisionality}\label{divcol}

\begin{figure}
	\centering
		\includegraphics[width=\linewidth]{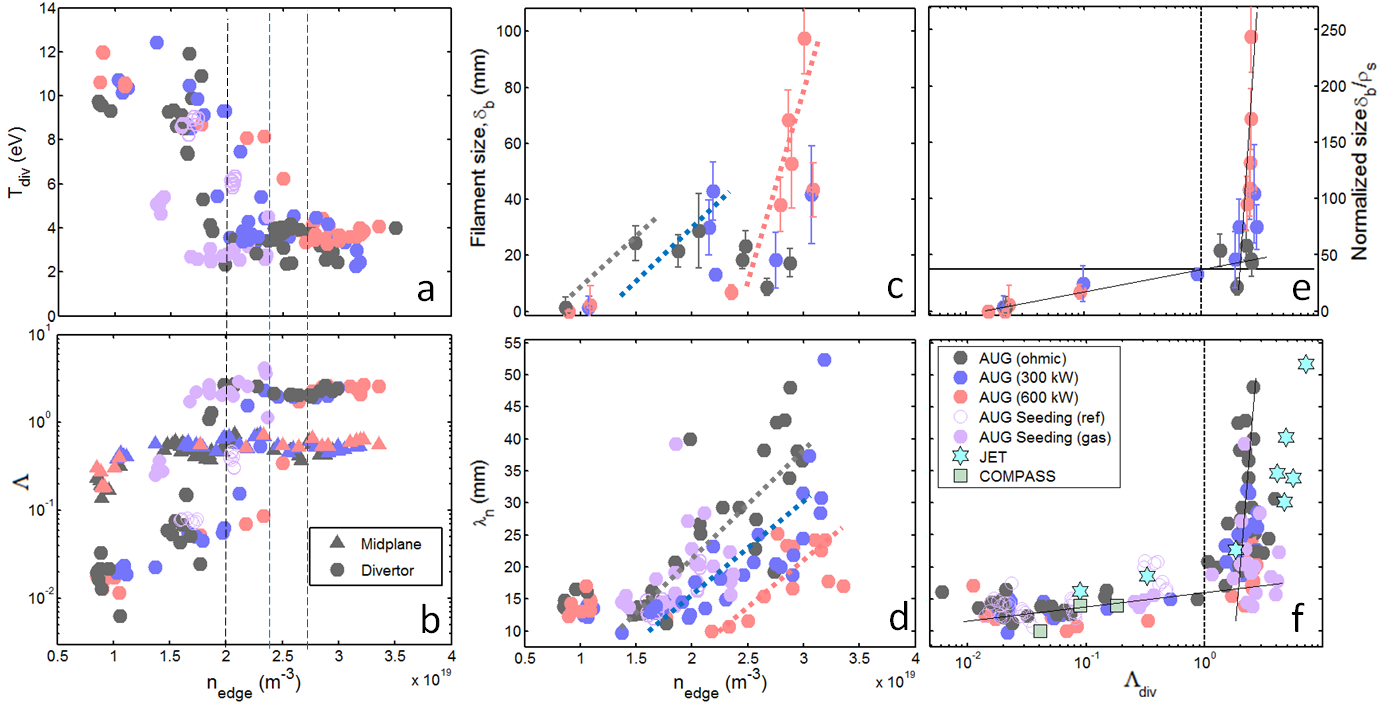}
	\caption{\textit{L-mode shoulder formation experiments. Color indicates heating power and seeding. Circles/stars/squares indicate data from AUG/JET/COMPASS; a) Evolution of divertor $T_e$ with plasma density.  b) Collisionality at the divertor/midplane (circles/triangles) as a function of density. Vertical dashed lines indicate the onset of detachment at the LFS divertor; c) and d) Evolution of filament perpendicular size and $\lambda_n$ as a function of edge density; e) and f) Evolution of filament size and $\lambda_n$ as a function of divertor collisionality.}}
	\label{fig:1}
\end{figure}

In order to assess the importance of midplane and divertor collisionalities in the shoulder formation, a series of L-mode density ramps were carried out on AUG\cite{Carralero15}. A lower single null, edge optimized configuration ($B_T = 2.5$ T, $I_p = 800$ kA, $q_95 \simeq 5$) was selected to replicate previous experiments \cite{Carralero14}. Different heating powers (including pure ohmic, $300$ and $600$ kW of ECH power) were used. Also, some of the $300$ kW discharges using nitrogen seeding were included in the analysis (although no filament data is available in these). Density profiles were measured using a lithium beam \cite{Fischer08, Willendorfer13}, and filaments in the far SOL ($\rho \simeq 1.02$) were characterized using a multipin Langmuir probe mounted on a midplane manipulator (MPM) \cite{Carralero14}. Also, divertor density and electron temperature were measured by flushed mounted Langmuir pins at the target plates. As can be seen in Fig. \ref{fig:1}a, the electron temperature, $T_e$, drop at the divertor associated to the high recycling regime of the LFS divertor takes place at different densities (indicated here as the line integrated density at the edge, n$_{edge}$, measured by interferometry) depending on power and seeding. These results, as seen in Fig. \ref{fig:1}b, in $\Lambda_{div} > 1$ being achieved at different n$_{edge}$ values. Instead, as will be discussed in section \ref{T}, $T_e$ does not change substantially in the midplane far SOL and $\Lambda_{mid}$ remains constant practically for the whole range of n$_{edge}$. In Figs. \ref{fig:1}c and \ref{fig:1}d, it is clear that the transition does not take place for a given density value, as filament perpendicular size, $\delta_b$, and the density e-folding length in the far SOL, $\lambda_n$, increase at different n$_{edge}$ values (in fact, in Fig. \ref{fig:1}c, the existence of a transition is only clear for the $600$ kW case). $\Lambda_{mid}$ is neither the driving parameter, as it  remains constant through the transition in all cases. Instead, as shown in Fig. \ref{fig:1}e and \ref{fig:1}f, all points converge into a single curve when $\delta_b$ and $\lambda_n$ are represented as a function of $\Lambda_{div}$ regardless of heating power or seeding. In this case, two clear regimes can be seen, with global particle transport strongly enhanced for $\Lambda_{div} > 1$. Interestingly, seeded discharges (represented in Fig. \ref{fig:1} as hollow/solid violet circles indicating the times in the discharge before/after the seeding) converge with the non-seeded ones, indicating that the transition is independent of how high divertor collisionality is achieved. In Fig. \ref{fig:1}f data points from JET and COMPASS have also been included: in JET, where a similar range of $\Lambda_{div}$ values was covered, a transition remarkably similar to the one at AUG is observed\cite{Carralero15}. In COMPASS, collisionality in the divertor has been recently measured \cite{Cavalier16} during L-mode density ramps. Following the same trend, since $\Lambda_{div} < 1$ for the whole data set due to a more limited range of achievable densities, all data points are in the low collisionality branch of the figure.\\

\section{A change of filament scaling}\label{scal}

\begin{figure}
	\centering
		\includegraphics[width=0.5\linewidth]{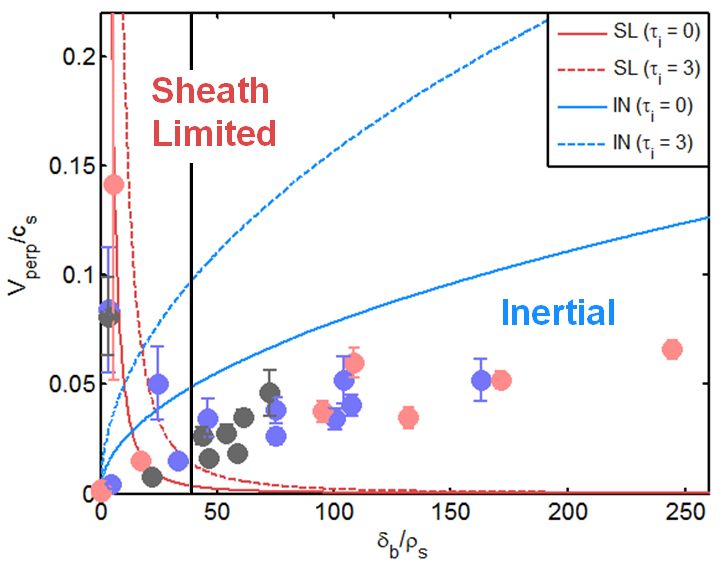}
	\caption{\textit{Filament scaling. Color code as in Fig. \ref{fig:1}. Black solid line indicates the regime separation, as observed in Fig. \ref{fig:1}e. Red and black lines indicate IN and SL regimes, as described in \cite{Carralero15}. Solid or dashed line indicates $\tau_i = 0$ and $\tau_i = 3$, respectively.}}
	\label{fig:2}
\end{figure}

As explained in the introduction, a transition between SL and IN filamentary regimes has been proposed as the explanation for the shoulder formation \cite{Dippolito06}. This would be consistent with the strong change of scaling of filament size shown in Fig. \ref{fig:1}e, which clearly mirrors that of $\lambda_n$. Such a transition had been observed in basic plasmas \cite{Theiler09}  but was yet to be confirmed in fusion-relevant machines like AUG. Therefore, in order to verify this hypothesis, the scaling of filament size as a function of perpendicular velocity was studied in the experiments described in the previous section\cite{Carralero15}. The velocity of sheath limited filaments scale as $v_\bot \propto (\delta_b)^{-2}$ \cite{Krash07}, while the velocity of filaments in the inertial regime scale as $v_\bot \propto (\delta_b)^{1/2}$ \cite{Garcia06}. By using a multipin probe head to calculate correlations between radially and poloidally spaced pins, the sizes and perpendicular velocities of conditionally averaged filaments during the density ramps were obtained \cite{Carralero14}. As can be seen in Fig. \ref{fig:2}, a clear transition in the scaling is observed around $\Lambda_{div} = 1$, with filaments in the $\Lambda_{div} < 1$ branch scaling according to the SL  model, and filaments in the $\Lambda_{div} > 1$ branch according to the IN regime. The transition is indicated in the figure by a vertical solid line which corresponds to the size for which the collisionality threshold is achieved in Fig. \ref{fig:1}e, where it is marked as a horizontal solid line. To illustrate this, theoretical predictions \cite{Manz13} for both regimes have been also represented in the figure considering both cold and warm ions, with $\tau_i = T_i/T_e$. Interestingly, IN regime filaments are clearly more consistent with the cold ion approximation. These novel results confirm that, coinciding with the onset of the density shoulder, the disconnection of filaments caused by a critical collisionality value leads to a transition of the propagation mechanism of filaments, which become larger and denser.\\

\section{Effect of the shoulder formation on SOL temperatures}\label{T}

\begin{figure}
	\centering
		\includegraphics[width=0.8\linewidth]{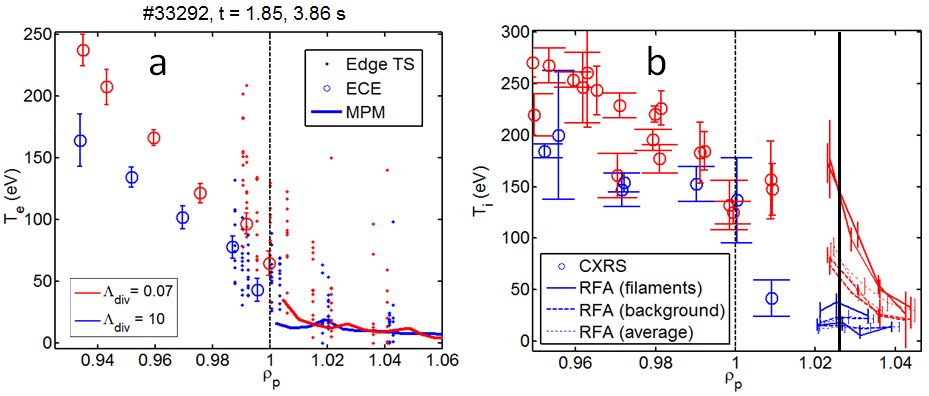}
	\caption{\textit{L-mode temperature measurements. Red/blue color indicates low/high collisionality; a) Radial profiles of $T_e$ , as measured by the Langmuir probe, edge Thomson Scattering and ECE; b) Radial profiles of $T_i$. Solid/dashed/dotted lines stand for filaments/background/average. Symbols correspond to CXRS measurements. Solid black line indicates the measurement position of the $E \times B$ analyzer experiments described in Fig. \ref{fig:4}.}}
	\label{fig:3}
\end{figure}

The evolution of $T_e$ and $T_i$ through the onset of the density shoulder has been investigated, as it determines the amount of energy advected by filaments. First, $T_e$ at the outer midplane (OMP) was measured by means of a swept Langmuir pin installed on the MPM. By plunging the probe close to the separatrix at high and low $\Lambda_{div}$ conditions, a typical radial profile can be obtained for each case. As can be seen in Fig. \ref{fig:3}a, the electron temperature profiles are roughly insensitive to $\Lambda$ in the far SOL, where $T_e$ remains roughly constant at $T_e \simeq 10-15$ eV. This is consistent with previous measurements in AUG \cite{Nold10}, and with the generally accepted picture of electrons losing most of their energy by parallel conduction within the first mm radially outside of the separatrix \cite{Stangeby}. At high collisionality, a drop in $T_e$ is observed in the near SOL instead. This can simply be explained as the result of pressure conservation as the density is increased to achieve high $\Lambda_{div}$ while the heating power is kept constant. \\

Second, $T_i$ has been measured in the far SOL by means of a Retarding Field Analyzer (RFA)\cite{KocanRFA} installed on the MPM \cite{CarraleroEPS15}. In this case, discharges with different constant density levels were carried out on AUG featuring low and high collisionalities ($\Lambda_{div} \simeq 0.01$ and $\Lambda_{div} \simeq 10$). In each case, the MPM was plunged to different distances to the separatrix in order to obtain radial ion temperature profiles in the far SOL. Unfortunately, in this case no $T_i$ measurements are available in the near SOL. Using the ion saturation current ($I_{sat}$) measured in the entrance slit of the RFA to carry out a conditional average, $T_i$ values of filaments and background were calculated separately. This technique is explained at length in \cite{KocanRFA}. As can be seen in Fig. \ref{fig:3}b, a strong reduction of $T_i$ takes place after the shoulder is formed: for $\Lambda_{div} \simeq 0.01$, a radial exponential decay can be seen both for the filaments and the background, while for $\Lambda_{div} \simeq 10$ the profile seems to remain radially constant at $T_i \simeq 20$ eV for both. If an average e-folding length is fitted to the $T_i$ profile starting at the separatrix, it goes from $\lambda_{Ti} \simeq 30 mm$ below the transition, to $\lambda_{Ti} \simeq 8 $mm above it. This result is in good agreement with the fact that disconnected filaments fit better with the cold IN regime, as discussed in section \ref{scal}. Before the transition, filaments are clearly warmer than the background at the innermost positions, while similar values are measured for high $\Lambda_{div}$. Interestingly, the filament $T_i$ value, which seems to increase exponentially as the separatrix is approached, is already similar to $T_{i,sep}$ at $\rho \simeq 1.02$ for the low $\Lambda_{div}$ case, suggesting that these structures must be generated inside the confined plasma, where $T_i$ is equal or greater than the one measured in the filament at the far SOL. In the high $\Lambda_{div}$ case, this argument can not be applied, as filaments are no longer hotter than the background. Similar results, including the reduction of $T_i$ at high densities and a $T_{i,fil}/T_{i,back} \simeq 3$ ratio at low densities, have been reported from MAST \cite{Allan16}. The reason for this cooling is not obvious since, as will be discussed in section \ref{disc}, pressure conservation is not sufficient to explain such a strong variation in $T_i$.\\

Last, the evolution of the $T_i$ distribution function has been observed by means of an $E \times B$ analyzer installed in the MPM. This diagnostic consists of a cavity opened to the plasma through a narrow slit aligned with the main magnetic field, similar to the one used in the RFA. In this case, a couple of electrodes inside the cavity create an electrostatic perpendicular field. This gives rise to an $E \times B$ drift affecting the ions entering the cavity, which deviate in the binormal direction a distance directly proportional to their time of flight in it (ie., inversely proportional to their initial parallel velocity). This effect is measured by a series of segmented collectors at the back of the cavity, which allow for a reconstruction of the $T_i$ with a time resolution only limited by the data sampling rate ($2$ MHz in the case of experiments carried out at AUG). A more detailed description of the $E \times B$ analyzer and the analysis details can be found in \cite{matthews84,KommExB}.\\

\begin{figure}
	\centering
		\includegraphics[width=0.5\linewidth]{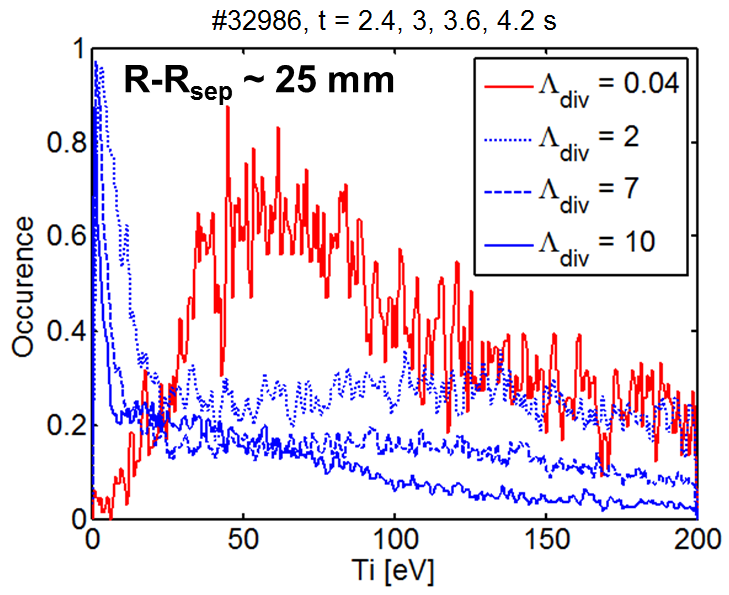}
	\caption{\textit{$T_i$ histogram from $E \times B$ measurements. The probability of a given $T_i$ is given for different $\Lambda_{div}$ values. }}
	\label{fig:4}
\end{figure}

In the $E \times B$ experiments, a density ramp was carried out at constant heating power while the MPM was plunged several times to $R-R_{sep} = 25$ mm. To allow for comparison with RFA measurements, this position is indicated as a solid black line in Fig. \ref{fig:3}. The plunge duration was set to $150$ ms in order to collect enough data to construct histograms of the $T_i$ measurements. As shown in Fig. \ref{fig:4}, the PDF of the $T_i$ displays a substantial change over the $\Lambda_{div} = 1$ transition: For $\Lambda_{div} < 1$, the distribution resembles a Gaussian with a strong positive tail. The center of the Gaussian corresponds to $T_i \simeq 60$ eV, while the tail corresponds to values in the $T_i = 100-200$ eV range. These values correspond with those measured in the same position with the RFA for background and filaments, respectively. For $\Lambda_{div} > 1$, the Gaussian shape disappears and a large population of cold ions appears in the $T_i < 5$ eV range. The hot ion tail remains unchanged. However, as the collisionality increases, the cold ion population eventually dominates the distribution function. Again, this is consistent with the measurements for $\Lambda_{div} > 1$ shown in Fig. \ref{fig:3}: due to its much lower temporal resolution, the RFA averages $T_i$ over the PDF, thus showing the increase of the cold ion population as drop of the $T_i$. The measurement of $T_i$ is limited by the spatial resolution of the segmented collector, so $T_i > 200$ eV could not be resolved. Therefore, the maximum temperature achieved by ions in the tail is undetermined.\\

\section{H-mode shoulder formation}\label{H}

\begin{figure}
	\centering
		\includegraphics[width=\linewidth]{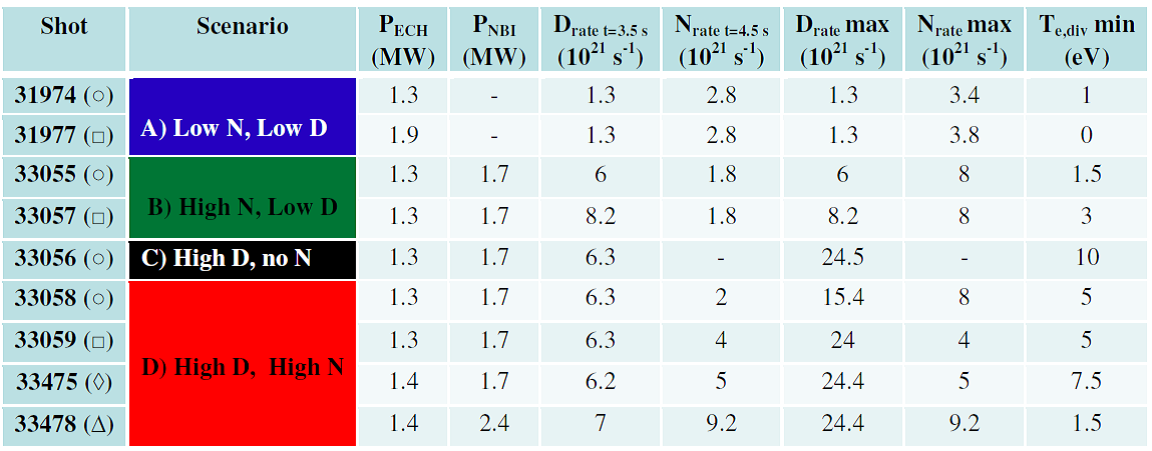}
	\caption{\textit{Main parameters of the H-mode experiments \cite{Carralero16}. The four scenarios A-D are grouped in different colors (respectively, blue/green/black/red), based on their $D_{rate}$, $N_{rate}$ values.}}
	\label{fig:5a}
\end{figure}

Since next generation machines are foreseen to operate in H-mode, any attempt to predict their behavior must be valid in this regime of confinement. Previous studies at low densities show no drastic change of filamentary transport in L- and H-mode plasmas \cite{Fuchert14}. Nevertheless, an additional line of work was dedicated to determine if the findings in L-mode plasmas also apply to H-mode plasmas \cite{Carralero16}. With this aim, a series of discharges has been carried out in AUG in which the L-mode scenario with 300 kW of heating power used in previous work was achieved as a reference, and then brought into H-mode by increasing the heating power up to 1-4 MW. Then, the collisionality in the divertor was raised by increased fueling or nitrogen seeding, in order to reach the $\Lambda_{div} > 1$ condition. In order to disentangle the effects of nitrogen and deuterium fueling on the $\Lambda_{div}$ values, different fueling rates for both gases ($N_{rate}$ and $D_{rate}$) were used, roughly dividing the data set in four scenarios, as detailed in Fig. \ref{fig:5a}: A) low power discharges with low $N_{rate}$ and $D_{rate}$ values; B) discharges including both NBI and ECH heating, strong nitrogen seeding and a low $D_{rate}$; C) A discharge in which $\Lambda_{div} > 1$ is achieved only by means of a strong density fueling with no nitrogen; D) discharges with full power, and both high $N_{rate}$ and $D_{rate}$ values. As before, density e-folding length, $\lambda_n$, and filament size, $\delta_b$, were measured with the LiB and MPM diagnostics, respectively. Thermoelectric currents to the divertor \cite{Kallenbach10}, $I_{div}$, are used to detect ELMs in order to separate conditionally averaged inter-ELM values of all measurements.\\

\begin{figure}
	\centering
		\includegraphics[width=0.8\linewidth]{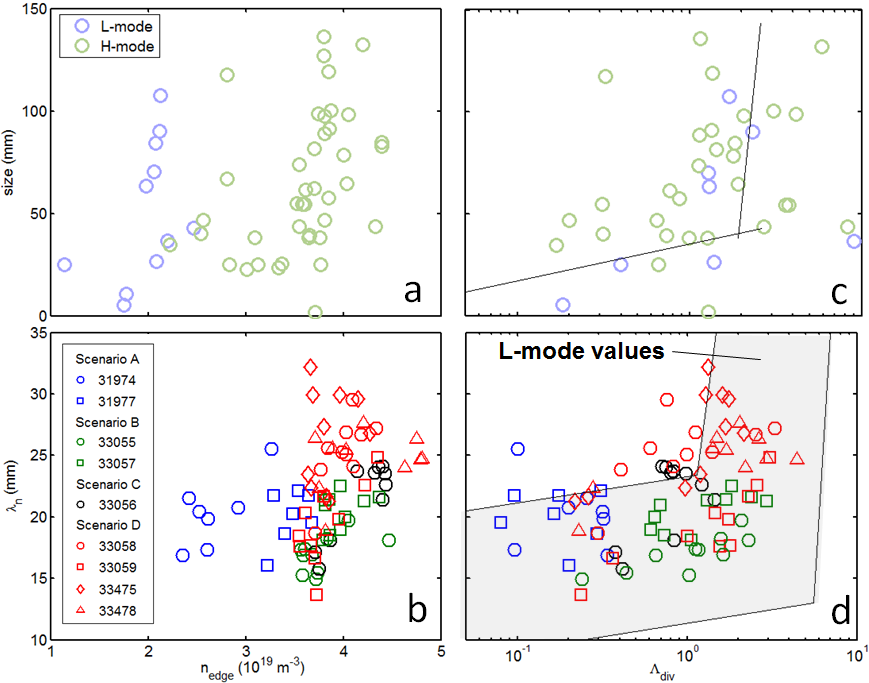}
	\caption{\textit{H-mode shoulder formation experiments \cite{Carralero16}; Size of filaments as a function of a) density and c) collisionality. Blue/Green colors stand for L/H-mode. Solid lines in c) correspond to the L-mode scaling shown in Fig. 1e); b) and d) show respectively the e-folding length as a function of density and collisionality. Color stands for the experimental scenario. Shaded area in d) represents the range of L-mode values shown in Fig. 1f).}}
	\label{fig:5b}
\end{figure}

Results are presented in Fig. \ref{fig:5b} following the same arrangement employed in Fig. \ref{fig:1} for L-mode results: First, in Fig. \ref{fig:5b}a and c, $\delta_b$ is presented as a function of n$_{edge}$ and $\Lambda_{div}$, respectively. As can be seen, a filament transition takes place for L-mode filaments at $n_{edge} \simeq 2\cdot10^{19}$ m$^{-3}$, which coincides with the value for the $300$ kW discharges in Fig \ref{fig:1}. When additional power is injected to access H-mode, the collisionality is reduced again and $\delta_b$ drops. Finally, when density is further increased over a second threshold, $n_{edge} \simeq 3.5\cdot10^{19}$ m$^{-3}$, a similar transition takes place for H-mode inter-ELM filaments. As in Fig. \ref{fig:1}e, both transitions converge and take place around $\Lambda_{div} \simeq 1$ when represented as a function of collisionality. The onset of the shoulder follows the same general trend -albeit not as pronounced- as in L-mode when considering the point cluster as a whole (represented in Fig. \ref{fig:5b}b and d): shoulder formation coincides again with both the $n_{edge} \simeq 3.5\cdot10^{19}$ m$^{-3}$ threshold for H-mode filament transition and with the general $\Lambda_{div} \simeq 1$ transition condition, with points taking similar values as the ones in Fig. \ref{fig:1}f (indicated in Fig. \ref{fig:5b}d as a shaded area). However, when the four scenarios are considered individually, a clearer trend is found: discharges from scenario A display low levels of $\Lambda_{div}$ and do not access the higher transport regime. The same is valid essentially for scenario C, albeit slightly higher $\Lambda_{div}$ (and thus $\lambda_n$) values are achieved. Instead, all discharges from scenario D develop a clear shoulder, achieving a significant increase in $\lambda_n$ at higher values of $\Lambda_{div}$. Interestingly, discharges in scenario B, displaying similar $\Lambda_{div}$ and n$_{edge}$ values as those in scenario D, fail to achieve high $\lambda_n$ values. This reveals a more complex picture than in L-mode, as $\Lambda_{div} > 1$ seems to be necessary but not sufficient for the shoulder formation, with the level of $D_{rate}$ playing also some role in it. Since scenarios B and D feature similar n$_{edge}$ values, such a role is not likely to be related to the fueling of the main plasma, but instead to some mechanism taking place in the SOL. \\

It must be pointed out that due to limitations in the data set, the same individual discussion can not be made with the filaments: while the general trend seen in Fig. \ref{fig:5b}c indicates that filaments undergo the transition at $\Lambda_{div} =1$ very much like they do in L-mode, insufficient measurements are available to discriminate the behavior of the different scenarios. Therefore, although there is a general correlation between transport and filament size ($\lambda_n \propto \delta_b$ approximately) and therefore filaments are generally not as large in scenario B as in scenario D, it is not possible to state with certainty if the lack of shoulder in scenario B is also caused by a lack of filamentary transition, or if it happens despite such transition being achieved \cite{Carralero16}. The relevance of this matter will be further discussed in section \ref{disc}.\\

\section{EMC3-EIRENE Simulations}\label{sim}

The problems arising from the results presented in the last two sections, namely the determination of the mechanism behind the ion cooling after the shoulder formation, and the determination the role of $D_{rate}$ as a secondary threshold in H-mode experiments may involve atomic physics (such as ionization and CX collisions) which would require a discussion of the neutral density, $n_{neutral}$ and flux profiles. Since this information can currently not be obtained from diagnostics in AUG with sufficient spatial resolution, we have resorted to simulate the conditions of the SOL before and after the transition with a numerical transport model. \\

In a previous work \cite{Lunt15}, two AUG L-mode discharges featuring $\Lambda_{div}<1$ and $\Lambda_{div}>1$ were simulated using the Edge Monte Carlo 3D-EIRENE (EMC3-EIRENE) code package. This code relies on the detailed atomic physics model developed for EIRENE \cite{Reiter05} and is particularly well suited for this problem as it can provide a full 3D treatment of transport, including the effect of non-toroidally symmetric plasma facing components (PFC) such as limiters, antennas, etc., where most of the plasma-surface interaction with the main chamber wall takes place. Also, this code does not require its grid to be aligned with the flux surfaces, allowing for a simple extension beyond the second separatrix. In \cite{Lunt15}, it was demonstrated that, in order to match the experimental density profiles, it was required to enhance transport in the SOL by increasing $D_{\bot}$ by an order of magnitude. This can be regarded as a diffusive approximation to the effect of the filamentary transport observed in experiments \cite{Carralero14}, and is consistent with the increase of anomalous diffusion associated with shoulder formation measured in DIII-D and Alcator C-mod (so strong that brought its diffusive nature into question) \cite{Lipschultz05}. It was observed that the midplane density profiles could not be reproduced in the $\Lambda_{div}>1$ case if the PFC were removed from the simulation, indicating that the interaction with the main wall plays a role in the formation of the shoulder.\\

\begin{figure}
	\centering
		\includegraphics[width=\linewidth]{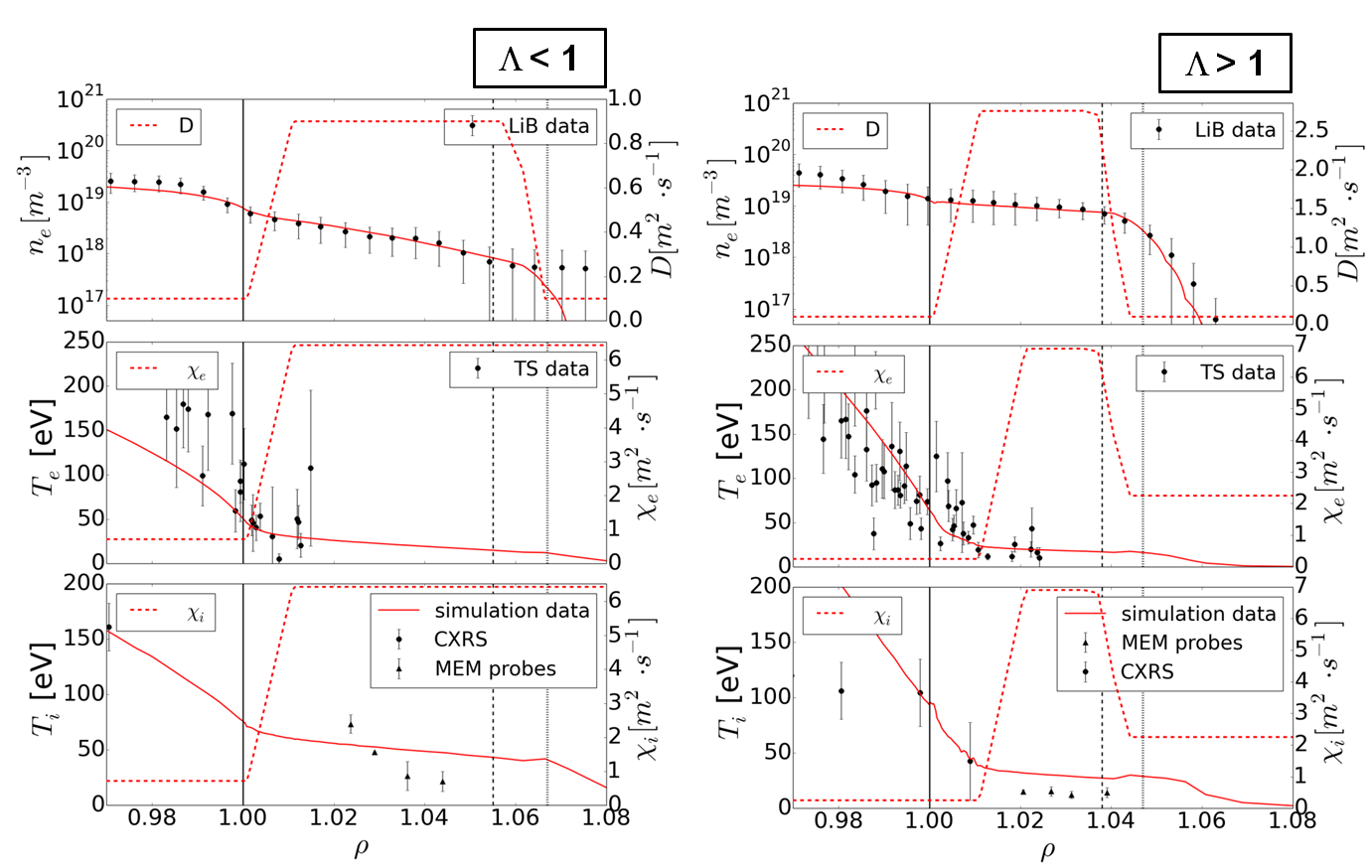}
	\caption{\textit{EMC3-EIRENE simulations for the $\Lambda_{div} < 1$ and $\Lambda_{div} >1$ cases, corresponding to cases A and B in \cite{Lunt15}. From top to bottom, experimental and simulated $n_e$, $T_e$ and $T_i$ profiles are represented by black symbols and solid red lines, respectively. Dashed lines represent the corresponding diffusive transport coefficients $D_{\bot}$,$\chi_e$,$\chi_i$ assumed in the simulation. Vertical gray dashed/dotted lines indicate the inner heat shield/OMP limiter surfaces.}}
	\label{fig:6}
	\end{figure}

Starting from the results in \cite{Lunt15}, input parameters of the simulation were adjusted to match not only the density, but also the $T_e$ and $T_i$ SOL profiles in the midplane and improve the agreement with X-point and divertor data. This improvement was partly achieved by assuming a poloidal dependence of the transport coefficient, $D_{\bot}$ with a maximum at the OMP, as would be expected from filamentary transport\cite{Krash01}. A similar method has recently been found to be required to simulate the formation of the HFS high density front using SOLPS \cite{Reimold16}. Also, the enhanced $D_{\bot}$ associated with the filaments is removed after the limiter shadow is reached, in good agreement with recent experimental results \cite{Fuchert16}, which show how filaments eventually revert to SL regime when their connection length is reduced sufficiently. The results of this simulations, shown in Fig. \ref{fig:6}, reveal good agreement with available $n_e$, $T_e$ and $T_i$ data at the OMP for both the low and high density case. These simulations and their results are explained in more detail in \cite{DIsa16}, where reasonable agreement with $n_e$ and $T_e$ at the X-point and divertor regions is also shown.\\

\begin{figure}
	\centering
		\includegraphics[width=.8\linewidth]{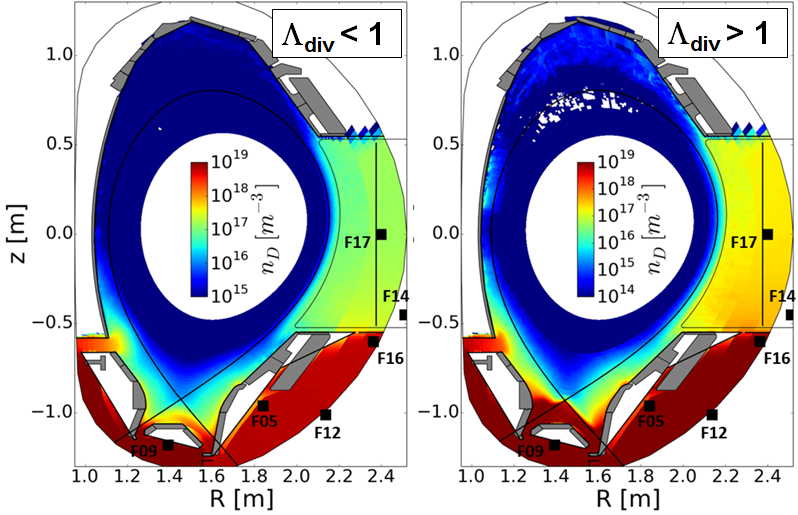}
	\caption{\textit{EMC3-EIRENE results. Poloidal section of the atomic Deuterium neutral density, as calculated by the code for the $\Lambda_{div} < 1$ and $\Lambda_{div} > 1$ cases. The positions of several neutral gauges are indicated by black squares.}}
	\label{fig:7}
	\end{figure}

\begin{figure}
	\centering
		\includegraphics[width=.5\linewidth]{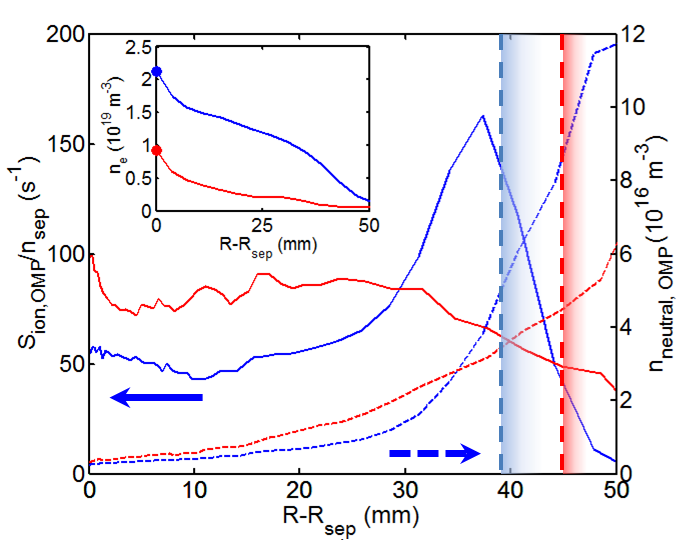}
	\caption{\textit{EMC3-EIRENE results in the OMP. Red/blue color indicates low/high collisionality. Neutral density profiles are displayed as dashed lines. Ionization rates are normalized to their maximum value and displayed as solid lines. The OMP limiter position is indicated in each case by a colored vertical dashed line. The insert shows the density profiles in the same region, with the value used for normalization highlighted with a circular marker.}}
	\label{fig:7b}
	\end{figure}

Once the available measurements have been matched by the code, a reasonable approximation to the real average neutral distribution can be obtained from its output: In Fig. \ref{fig:7}, a poloidal cross-section of the $n_{neutral}$ is shown for the two cases with $\Lambda_{div}$ below and above unity. Although no direct measurement of $n_{neutral}$ is available in the SOL, the results of the simulation can be compared with neutral fluxes recorded by gauges placed in the chamber wall, also displayed in Fig. \ref{fig:7}. For $\Lambda_{div} < 1$, the measurements of the two gauges in the midplane (F17 and F14) match the simulation results within 30\% error bars. However, in the high density case, the simulation seems to underestimate the measurement of the $n_{neutral}$ by a factor $3-4$\cite{DIsa16}. Since plasma-neutral reaction rates are strongly non-linear functions, the worse agreement in the high $\Lambda_{div}$ case could be the result of mean field model being insufficient to describe a SOL dominated by strong turbulence \cite{Thry}. The first $50$ mm  of the SOL of the simulated neutral profile in the OMP are shown for both cases in Fig. \ref{fig:7b}. It can be seen that $n_{neutral}$ behind the OMP limiter shadow is higher in the $\Lambda_{div} > 1$. This can be seen as well in Fig. \ref{fig:7}. However, it can also be seen how this situation reverses as the separatrix is approached, with the high collisionality $n_{neutral}$ falling below the low collisionality one around $R-R_{sep} \simeq 35$ mm.

	This faster decay of the neutral population coincides with the appearance of a strong ionization front in front of the limiter. This can be seen in Fig. \ref{fig:7b}, where the ionization rates calculated by the code for both cases have been represented normalized to the plasma density at the separatrix to account for the increase of the ionization merely related to the general density increase after the shoulder (which is represented in the insert of the figure). As can be seen, in the low collisionality case, the ionization is distributed across the SOL with its maximum value at the separatrix. In this case, the limiter does not seem to play a relevant role neither in the ionization nor in the $n_{neutral}$ profile. Instead, when the shoulder is formed, a sharp peak appears in front of the limiter (which marks the point in which the plasma density declines, as can be seen in Fig. \ref{fig:6}, right) and then falls substantially for the rest of the SOL.

	\section{Discussion}\label{disc}
	
	Summarizing the results presented so far, three main phenomena have been observed associated to the shoulder formation, described here as a critical increase of $\lambda_n$ across the SOL:
	\begin{enumerate}
	\item [-] First, as shown in Fig. \ref{fig:1}, when the critical value of $\Lambda_{div}$ is achieved, filament size increases substantially, as well as perpendicular transport of particles associated to filaments in L-mode.
	\item [-] Second, as shown in Figs. \ref{fig:3} and  \ref{fig:4}, when the shoulder forms, $T_i$ drops across the SOL and filaments thermalize with the background. This cooling seems to be related to the appearance of a large population of cold ions which coexist with the hot tails observed at low collisionalities.
	\item [-] Last, as shown in Fig. \ref{fig:7b}, EMC3-EIRENE simulations indicate that, under the SOL conditions achieved after the shoulder formation, ionization tends to concentrate in front of the OMP limiter and neutral density is reduced in the far SOL.
	\end{enumerate}

The first point has been the subject of most of previous work \cite{Carralero14,Carralero15}, and seems to be explained as the result of the filamentary transition resulting from electrical disconnection from the target. However, the filament transport transition alone can not account for the second and third effects: Indeed, as can be seen in Fig. \ref{fig:3}, the moderate reduction in the edge $T_i$ at the separatrix associated with the density increase with constant power can not be accounted for this change, as it goes from $T_{i,sep} \simeq 175$ eV to $T_{i,sep} \simeq 150$ eV across the transition. Instead, filament and background temperatures drop by a factor 8 and 4, respectively. Another possibility would be thermalization between the ion and electron species, which would allow electrons to effectively remove this energy from the OMP via parallel conduction. The loss of ion energy via this channel can be calculated as \cite{Wesson}:

\begin{equation}
\frac{3}{2}n\frac{\partial T_i}{\partial t} = Q_i,
\end{equation}

where

\begin{equation}
Q_i = 3\frac{m_e}{m_i}\frac{n}{\tau_{ei}}(T_e-T_i)
\end{equation}

is the energy transfer rate from the ions to the electrons and $\tau_{ei} \simeq 6.4 \times 10^{14} T_e^{3/2}/n$ is the electron-ion collision time (with $T_e$ expressed in keV). Therefore, for $T_i>T_e$, the ion-electron thermalization time can be approximated as

\begin{equation}
\tau_{t,ie} \simeq \frac{1}{2}\frac{m_i}{m_e}\frac{T_i}{T_i-T_e}\tau_{ei}.
\end{equation}

According to the data in Fig. \ref{fig:3}, the cooling of the filaments is already complete $25$ mm in front of the separatrix. Therefore, the relevant values of $n$, $T_e$ and $T_i$ for an order of magnitude estimation of $\tau_{t,ie}$ can be obtained taking their average in the $R-R_{sep} \in [0,25]$ region. Taking $n \simeq 4.5 \times 10^{18}$ m$^{-3}$, $T_e \simeq 35$ eV and $T_i \simeq 175$ eV for the low collisionality case, $\tau_{t,ie} \simeq 2$ ms is obtained. This can be compared with the typical time a filament requires to travel through the same region, $\tau_\bot$. Taking a typical radial velocity range of $v_r \simeq 200-400$ m/s in the same region \cite{Carralero14}, $\tau_\bot \simeq 62-125 \mu$s values are obtained, clearly below $\tau_{t,ie}$. Therefore, before the shoulder is formed, ions and electrons are thermally decoupled. This result is not surprising given the strong temperature difference between the two species. Taking average values corresponding to this SOL region after the shoulder formation ($n \simeq 1.5 \times 10^{19}$ m$^{-3}$, $T_e \simeq 25$ eV and $T_i \simeq 90$ eV), a shorter thermalization time $\tau_{t,ie} \simeq 360 \mu$s is obtained as a result of the increased collisionality. Typical filament velocities after the shoulder formation are $v_r \simeq 400-800$ m/s  \cite{Carralero14}, resulting in $\tau_\bot \simeq 30-62 \mu$s. While this value is still smaller than $\tau_{t,ie}$ by at least a factor 5, the difference is probably not conclusive for this kind of order-of-magnitude analysis. This is aggravated by the uncertainties of $T_e$ measurements in the near SOL, where a difference of a few eV is determinant. A proper calculation of this effect goes beyond the scope of the present work, so the possibility that electrons evacuate a part of the ion energy can not be ruled out at this point. Nevertheless, the $E \times B$ analyzer measurements shown in Fig. \ref{fig:4} suggest that this mechanism can not play a dominant role: indeed, the low $T_i$ distribution observed after the shoulder peaks at values under $T_i = 5$ eV, which is substantially below $T_e$ at the same region. Therefore, this low temperatures can not possibly be the result of electron-ion thermalization. Besides, as discussed in section \ref{T}, the $T_i$ becomes less Gaussian-like after the transition, which is not the likely effect of a collisional process: should the electrons cool the ions, a more Gaussian distribution, centered around the electron temperature $T_e \simeq 15$ eV would be expected instead.\\

From these $E \times B$ analyzer results, a more likely explanation for this phenomenon can be inferred: the hot ions originated in the separatrix would not be cooling down, but joined by a population of cold ions being generated from the far-SOL neutrals, either by ionization or by CX collisions. This explanation would be in good agreement with the fact that the peak of the cold ion distribution is below $T_i \simeq 5$ eV, which is close to the Frank-Condon energies for dissociation ($\simeq 3$ eV) and dissociative ionization ($\simeq 5$ eV)  of $D_2$ molecules \cite{Sharp71}. It would also explain why the $T_i$ of filaments drops to that of the background, as ionization and CX reactions would be more likely in filaments -where the density is higher- than in the background, making the distribution of cold ions approximately proportional to the local electron density. In this situation, the conditional averaging used by the RFA (based on the $I_{sat} >2\sigma$ definition of filament, with $I_{sat}$ measured at the entrance slit \cite{KocanRFA}) would not be able to separate cold and hot ions. This is in contrast with the low $\Lambda_{div}$ situation, in which filaments transport mostly hot ions coming from the confined region, and the conditional averaging can effectively separate ions of different $T_i$.

\subsection{SOL neutral opacity}

	\begin{figure}
	\centering
		\includegraphics[width=.5\linewidth]{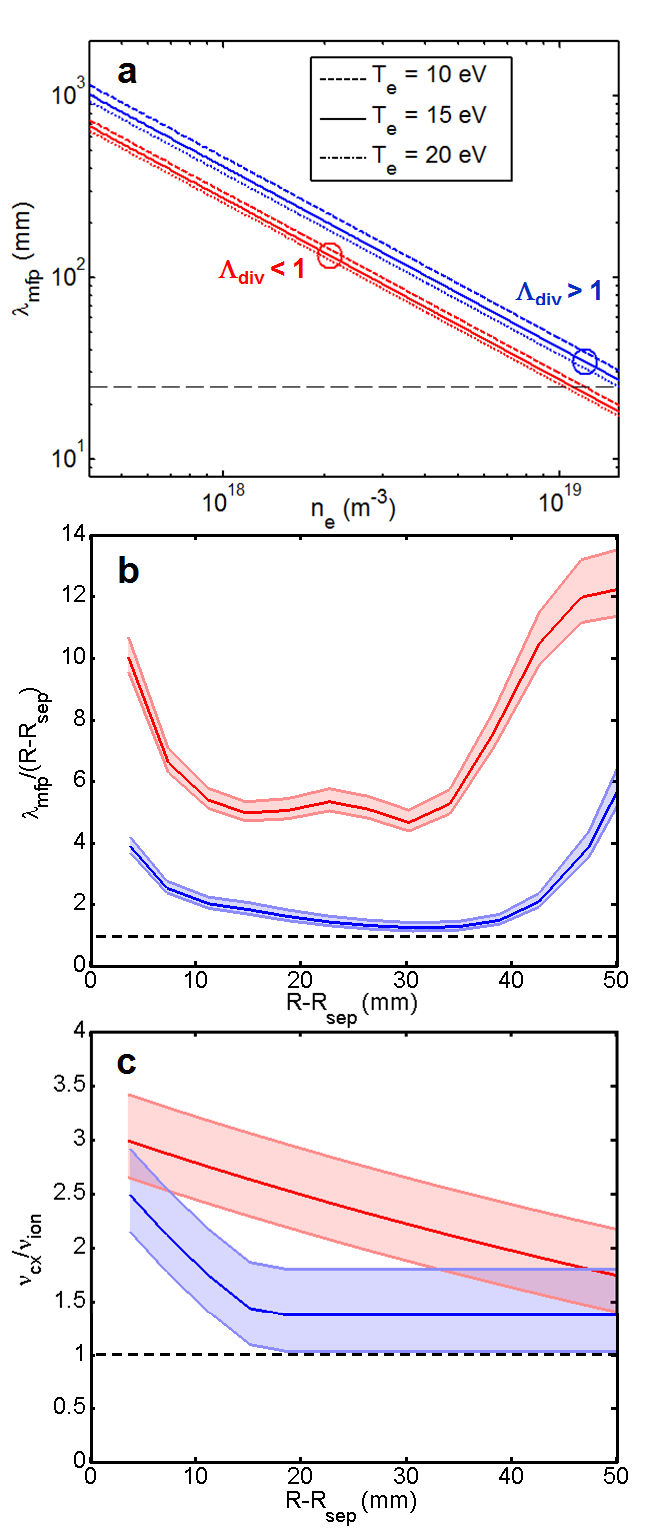}
	\caption{\textit{SOL opacity. a) Neutral mfp as a function of local density and electron temperature. Red/blue curves correspond to the $T_i$ values measured at $R-R_{sep} = 25$ mm for low/high $\Lambda_{div}$. Symbols indicate experimental $n_e$ values at the same position. Horizontal dashed line the distance to the separatrix. b) Radial profile of neutral mfp normalized to the local distance to the separatrix. Red/blue curves correspond to density profiles before/after the shoulder formation. Thick lines indicate typical $T_e = 15 eV$ case. Thin lines indicate $T_e = 10 eV$ and $T_e = 20 eV$ cases. Dashed line indicates $\lambda_{mfp}/(R-R_{sep}) =1$ c) Radial profile of CX vs. ionization rate ratio. Lines as in plot c. }}
	\label{fig:8b}
	\end{figure} 

While the emergence of a cold ion population provides a reasonable explanation for the observed cooling of the ions, it is still not clear why the neutrals affect the energy balance more effectively after the filament transition. A possible explanation for this has been proposed in the literature \cite{Lipschultz05}: the length of the mean free path (mfp) of the neutrals in the SOL may decrease substantially over the formation of the shoulder, leading to the onset of a ``recycling condition'', similar to the one found in the divertor. According to this idea, the plasma density increase in the far SOL would cause a rise in ionization and charge-exchange collision (CX) frequencies. This, in turn, would lead to a non-linear decrease of the neutral mfp, $\lambda_{mfp}$:
\begin{equation}
\lambda_{mfp} = \frac{v_n}{n_e(2\left\langle \sigma v\right\rangle_{CX} +\left\langle \sigma v\right\rangle_{ion})}, \label{eq:mfp}
\end{equation}
where the neutral velocity $v_n$ can be approximated using the above mentioned Franck-Condon dissociation energy and elastic collisions have been assumed to have approximately the same frequency as CX. This increase in the SOL opacity would prevent the neutrals from approaching the separatrix, increasing $n_{neutral}$ in the far-SOL and leading to enhanced ionization which would in turn cause an even further increase of density. This kind of feedback-loop could explain why the shoulder formation leads to the formation of a localized ionization region in front of the limiter and the reduction of neutral density close to the separatrix in the $\Lambda_{div} > 1$ case. Evidence for this process has been found in DIII-D and Alcator C-mod, where the formation of the shoulder has been related to the decrease of $\lambda_{mfp}/a$ \cite{Lipschultz05}. Similarly, shoulder formation in JET has recently been linked to ionization becoming dominant in the far SOL for high densities \cite{Wynn16}. Finally, such need for strong recycling would be consistent with the previously mentioned fact that shoulder formation in AUG can not be simulated with EMC3-EIRENE if the main wall is removed \cite{Lunt15}.\\

The evolution of the $\lambda_{mfp}$ in AUG with local $n_e$ at $R-R_{sep} = 25$ mm is displayed in Fig., \ref{fig:8b}a for the average $T_e = 15$ eV value displayed in Fig. \ref{fig:3}, along with the  $T_e = 10$ eV and  $T_e = 20$ eV cases. Typical $T_i = 75$ eV and $T_i = 20$ eV are respectively taken for the low/high collisionality cases (as seen in Fig. \ref{fig:3}b). As can be seen, before the shoulder formation, the $\lambda_{mfp}$ is substantially larger than the wall clearance, indicating that neutrals can freely travel across the SOL. Instead, when the shoulder is formed, $\lambda_{mfp}$ falls to values close to the width of the SOL, indicating that a large fraction of neutrals are no longer able to reach the separatrix. This result is refined in Fig. \ref{fig:8b}b by calculating the local $\lambda_{mfp}$ across the separatrix using typical density and $T_i$ profiles displayed respectively in Fig. \ref{fig:6} and \ref{fig:6}b and normalizing it by the local distance to the separatrix. Again, several values of $T_e$ are considered. As can be seen, the whole SOL is ``transparent'' to $5$ eV neutrals in the $\Lambda_{div} < 1$ case, while the ratio drops to almost 1 for a large part of the SOL for the $\Lambda_{div} >1$ case. Consistently, the region where the SOL becomes more opaque corresponds roughly to that in Fig. \ref{fig:7b} where the ionization peaks for the high $\Lambda_{div}$ case and the neutral density drops below the low collisionality case.\\

If this opacity threshold is also required in order to form the shoulder, ionization and CX reactions would tend to concentrate in a relatively narrow region in front of the first wall. This could lead to the emergence of a cold ion population like the one detected by the $E \times B$ analyzer, as opposed to the low $\Lambda_{div}$ situation, in which the effect of ionization and CX is not only lower, but also more spread across the SOL and the edge. CX and ionization rates are compared in Fig. \ref{fig:8b}c: it can be seen that $\nu_{CX} > \nu_{ion}$ in all cases for $\Lambda_{div} < 1$, indicating that most of the cold ions are generated in exchange for some of the hot previously present in the distribution. For $\Lambda_{div} > 1$, $\nu_{CX} > \nu_{ion}$ still holds, but the difference becomes smaller. This would be consistent with the strong reduction of the Maxwellian part of the pre-shoulder distribution (around $50-60$ eV) after the cold ion population appears. Why this range of energy is more affected than the $T_i > 100$ eV range is left for future investigation.\\

\begin{figure}
	\centering
		\includegraphics[width=0.8\linewidth]{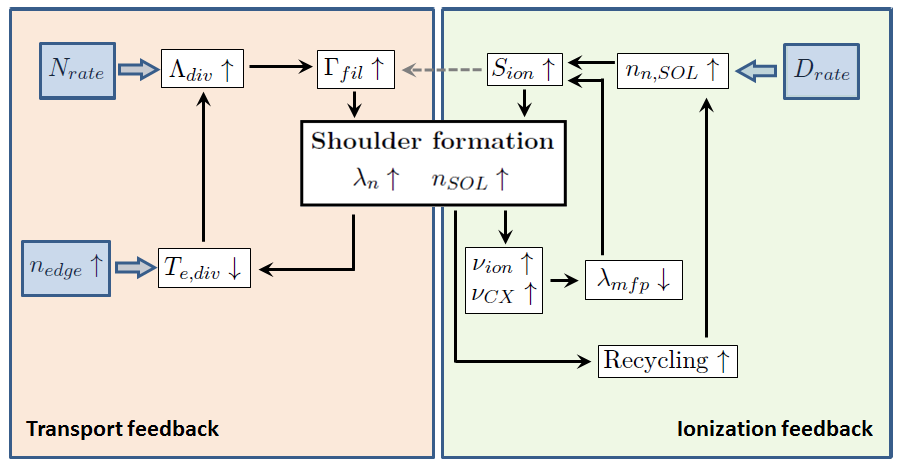}
	\caption{\textit{Summary of the mechanisms leading to the shoulder formation. Inputs are highlighted in blue. $n_{SOL}$ and $n_{n,SOL}$ stand respectively for the plasma and neutral densities in the far SOL.}}
	\label{fig:8c}
	\end{figure}

According to this view, summarized in Fig. \ref{fig:8c}, the shoulder formation would be the result of the combination of two feedback processes non-linearly related to the far-SOL density: first, as $n_{edge}$ rises, collisionality is strongly increased in the divertor region, eventually leading to the filament regime transition and the enhancement of filamentary transport. This contributes to the perpendicular spread of power, which leads again to an increase of $\Lambda_{div}$. This loop can be affected by impurity seeding $N_{rate}$, which cools the divertor and increases $\Lambda_{div}$. Second, the rise of density in the far-SOL leads to an increase of neutral opacity and main wall recycling, which strongly increases the neutral density in front of the main wall, enhancing the ionization source, $S_{ion} \simeq n_n v_n/\lambda_{mfp}$, and increasing the SOL density even further. In this case, deuterium fueling, $D_{rate}$, may act as a knob for the loop, increasing the neutral density in the far-SOL, $n_{n,SOL}$. This second loop could begin to explain the secondary D$_{rate}$ threshold discussed in section \ref{H}: In a situation where ELMs are flushing the SOL regularly and $\Lambda_{div}$ is only marginally below its critical value, a minimum fueling level might be required in order to maintain a critical level of neutral density in the far-SOL, capable of sustaining the ionization front. Also, this double condition could be explored to analyze other experiments where the $\Lambda > 1$ condition has been found insufficient to explain the shoulder formation, such as recent work at TCV \cite{Vianello16}, where a similar D$_{rate}$ has been identified, or experiments carried out in JET \cite{Wynn16b}, where the $\Lambda > 1$ condition triggers the shoulder with the horizontal divertor configuration (as in Fig. \ref{fig:1}), but not with the vertical configuration. The determination of the precise mechanism which could link $D_{rate}$ to midplane $n_{neutral}$ involves several problems which exceed the scope of the present study (such as the evolution of the divertor compression, the relevance of the bypasses behind the main chamber wall, etc.) and will be left for future work.\\

\subsection{Future possible scenarios}

\begin{figure}
	\centering
		\includegraphics[width=\linewidth]{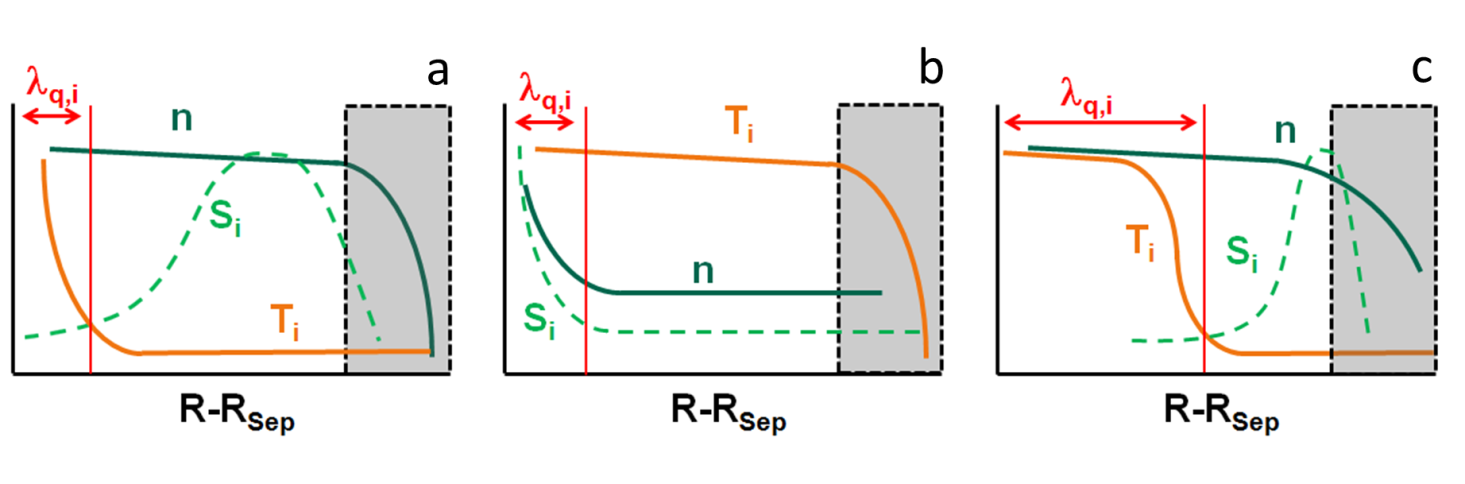}
	\caption{\textit{Possible future SOL scenarios. a) Standard scenario, with shoulder formation and strong $T_i$ decay. b) Unfavorable scenario, with no shoulder formation. c) Favorable scenario, with shoulder formation and $T_i$ decay region detached from the separatrix. }}
	\label{fig:8}
	\end{figure}

The main unresolved question left in the model presented in Fig. \ref{fig:8c} is the level of interconnection between the two loops and specially whether both are necessary in order to form the shoulder, or if it can be achieved with only one of them. In principle, it seems difficult to achieve a shoulder only via the ionization loop, as it is not started simply by an increase in $n_{edge}$ but requires the build up of density in the far SOL (or in other terms, a mechanism to transport the higher density at the separatrix towards the far-SOL is necessary, such as the filaments). Besides, enhanced heat convection may be required to transport enough electron energy to sustain high ionization rates in a region with very flat $T_e$ profiles \cite{Lipschultz05}. On the other direction, the increase of the ionization source contributes to the enhancement of filamentary transport by increasing the density in filaments. The minimum $D_{rate}$ value discussed in section \ref{H} suggests that both mechanisms are required. However, it must be taken into account that the $\Lambda_{div}$ was in those cases only marginally over the critical value, and ELM influence on the SOL was probably much larger than the one foreseen for ITER and specially DEMO \cite{Wenninger14}. Could then a sufficiently high $\Lambda_{div}$ lead to an enhancement of $\Gamma_{fil}$ strong enough to flatten the density profile without the contribution of local ionization? This question is not only academic: such link between the two mechanisms (indicated in Fig. \ref{fig:8c} with a gray, dashed arrow) could be decisive for the SOL structure of next generation devices. To illustrate this point, three different scenarios, summarized in Fig. \ref{fig:8}, are presented depending on the level of interaction between transport and ionization. It must be stressed here that the description of these scenarios is meant only for the sake of the conceptual discussion and not as an attempt to produce actual predictions on future machines.\\

In the first scenario, dubbed here ``standard scenario'', both loops would be required for the shoulder formation, but the ionization would extend far enough into the SOL to allow for the shoulder formation. This could be the result of a moderate clearance (such as the one to be found in ITER) or a sufficient neutral source in the midplane, such a strong recycling with a large fraction of neutrals reflected at high energies. This scenario would imply that the same behavior is found in ITER and DEMO as the one observed in present ITER stepladder experiments \cite{Carralero15}: since divertor collisionality will be substantially higher than in present day machines, the $\Lambda_{div} > 1$ condition will be fulfilled during standard operation. Therefore, according to this scenario, a shoulder will form leading to a substantial increase of $\lambda_n$, and the average $T_i$ will then display a sharp radial decay, as observed in Fig. \ref{fig:3}. In this scenario, shown in Fig. \ref{fig:8}a, the effect of density increase will likely be compensated by the drop in $T_i$ resulting in no net increase in the perpendicular heat transport to the wall. This has been already observed in AUG and JET \cite{Guillemaut16}. However, this means that the fraction of $P_{SOL}$ allocated to the ions will probably not be spread too far from the separatrix. Indeed, if the convection of energy is taken as the dominant transport mechanism, $q_\parallel \propto M_\parallel c_sn_e(T_e+T_i) \simeq n_e (T_e+T_i)^{3/2}$ can be defined, where $\nabla_\bot M_\parallel \ll \nabla_\bot q_\parallel$. Then, a parallel heat flux e-folding length can be obtained for the ions in the midplane, $\lambda_{q_i,\parallel} ^{-1} \simeq \lambda^{-1}_n+\frac{3}{2}\lambda_{T_i}^{-1}$ (since $\lambda^{-1}_{T_e} \ll \lambda^{-1}_{T_i}$ after the first few mm in the SOL). Taking AUG values from Figs. \ref{fig:1} and \ref{fig:3}, $\lambda_n \simeq 40$mm and $\lambda_{T_i} \simeq 8$ mm can be found after the shoulder formation, resulting in $\lambda_{q_i,\parallel} \simeq 5$ mm. This is not substantially higher than typical values for $\lambda_q$ calculated considering only electron conduction \cite{Sun15}, meaning that filamentary transport won't be able to introduce a substantial reduction of plasma heat loads in the divertor. Of course, as shown in Fig. \ref{fig:8b}, a fraction of the ion energy is transferred to the neutrals and scattered via CX collisions. How much of such power is deposited in the PFC, returned to the SOL plasma via ionization or recovered by the confined plasma by fueling is left as an open question.\\

In the second scenario, dubbed ``unfavorable scenario'' and shown in Fig. \ref{fig:8}b, both loops would also be required, but now the SOL would not be opaque enough for the second one to activate. Therefore, no shoulder would be formed. In next generation machines, neutral density far from the main wall might be substantially lower than in present day's, as clearance will be larger, separatrix temperatures and densities will be higher (resulting in lower $\lambda_{mfp}$), and neutral gas fueling will be substituted by pellet injection. As a result, neutral density and ionization rates could be substantially below current rates through the entire SOL. Therefore, if the increase of filamentary transport caused by $\Lambda_{div} > 1$ is not sufficient to form the shoulder without the contribution of a strong local ionization term, little to no increase of $\lambda_n$ would take place. This scenario could be similar to the H-mode scenarios without sufficient D$_{rate}$ discussed in section \ref{H}. In this case, ions would not cool down, potentially spreading high energy ions (like the ones seen in the hot tail of Fig. \ref{fig:4}) far away from the separatrix, which could have a serious impact on the first wall sputtering yield. This could be particularly serious in the case of ITER, where the clearance will not be substantially larger than that of AUG, and the use of Be as first wall material makes the prospect of having a non-negligible amount of 200 eV ions particularly concerning. In this case, no spread of the energy associated to the ions would take place either, as the large $\lambda_{T_i} \simeq 30$ mm will be compensated by the low $\lambda_n \simeq 15$mm, resulting in $\lambda_{q_i,\parallel} \simeq 8$ mm.\\

Finally, a last scenario can be defined if filaments can sustain a shoulder without ionization. This one, dubbed ``favorable scenario'' and displayed in Fig. \ref{fig:8}c, would be an intermediate case between the two first ones: In it, the enhancement of filamentary transport caused by disconnection would be able to sustain a shoulder without the need of local ionization and despite a high opacity and greater clearance in the SOL preventing the penetration of neutrals close to the separatrix. Pointing in this direction, recent experiments carried out in MAST show that the shoulder formation is possible without a strong ionization: In this large clearance machine, where no wall recycling is expected in the duration of a typical discharge \cite{Huang10}, the neutral density in the far SOL depends only on the edge plasma density, and no particular change in it is associated to flattening of the density profiles\cite{Millitelo16}. In this ``favorable scenario'', the far SOL would be thus divided into two parts: a hot layer close to the separatrix where neutral density is negligible, thus allowing for the conservation of high $T_i$ filaments, and a cold layer close to the wall where ionization and CX rapidly cool down the ion population and bring it to the low average $T_i$ values observed in the high $\Lambda_{div}$ in Fig. \ref{fig:3}. Since $\lambda_n$ would be high thanks to the transport of particles from the near-SOL, the recycling-ionization cycle could be sustained leading to a profile similar to the high $\Lambda_{div}$ scenario in Fig. \ref{fig:7b}. In this case, the main wall would be protected from hot ions like in the first scenario, but at the same time would feature a first region where both $\lambda_n$ and $\lambda_{T_i}$ would be high. This scenario can not be achieved in any present machine, as a larger clearance would probably be required as well as the possibility of retaining a hot SOL with a very collisional divertor. However, it is close to the standard operation conditions of DEMO and, to a lesser extent ITER (where, as already mentioned, clearance won't be much larger than in today's machines) \cite{Wenninger14,Federici14,Wenninger15} . If the $\lambda_n \simeq 40$ mm corresponding to the first scenario and the $\lambda_{T_i} \simeq 30$ mm corresponding to the second are taken, a $\lambda_{q_i,\parallel} \simeq 13$ mm could be achieved.\\

Understanding the mechanism determining the radial profile of $T_i$, as well as its relation to the shoulder formation and the penetration of neutrals in the SOL can be of paramount importance for the prediction of both sputtering of the main wall and divertor power loads. Regarding the main chamber sputtering in ITER, it must be stressed that the reduction of the mean $T_i$ by cold ion population does not eliminate the hot tails of the distribution: as can be seen in Fig. \ref{fig:4}, even in the highest $\Lambda_{div}$ case, a non-negligible amount of ions can be found in the $T_i > 100$ eV range, which would have a strong effect on a Beryllium wall. A dedicated study on the radial evolution of these tails under collisional conditions is required in order to decide if such hot ions actually travel to the wall or are eventually deflected by the increasing density of neutrals. In this sense, the question of whether cold ions are mostly generated by ionization or CX is relevant, as in the second case the reduction in hot ions could still mean the generation of a hot neutral population hitting the wall with energies of tens of eV. On the other hand, it must taken into account that only inter-ELM filamentary activity is being discussed here, while type-I ELMs are expected to dominate sputtering on the targets \cite{Guillemaut16b} and could therefore also play a relevant role in the main wall. The relative contributions to first wall erosion of intermittent, highly energetic type-I ELMs and the continuous, inter-ELM $200$-eV tail filaments described here is left as an open question for future work. In DEMO, sputtering in the main wall will likely be dominated by filamentary transport, as large ELMs will not acceptable for its operation. However, the first wall plasma facing material will be Tungsten \cite{Wenninger16}, meaning that the sputtering yield will be typically dominated by impurities rather than the direct impact of deuterium ions or neutrals, regardless of filament activity \cite{Birkenmeier15}. Although the impact of the potential arrival of filaments with $T_i$ exceeding the 200 eV can not be neglected, the impact of such filaments on impurities should be studied in order to have a complete idea of the problem. In particular, it would be important to know whether filaments may advect impurities as efficiently as deuterium, and whether these impurities can be expected to have equally high temperatures (eg., because they have been transported from the confined region).\\

Regarding the effect on the divertor loads, one conclusion of this work is that the energy carried by the ions must be taken into account when predicting heat loads, as there is no reason to assume that it will be substantially lower than the fraction corresponding to the electrons, and it may be spread over a much wider radial scale. In this sense, the $\lambda_{q_i}$ must not be confused with the $\lambda_q$ frequently used in the literature, which only considers the effect of electron parallel conduction in the OMP \cite{Sun15,Fundamensky05}. How this energy is transported into the divertor is a much more complicated issue, which includes all kinds of effects not mentioned here, such as thermalization with electrons (which will eventually ensue since $\nu_{ei}$ increases as the divertor is approached), flux expansion, X-point shear, radiation, etc. Nevertheless, it seems reasonable to assume that none of these processes will reverse the spread of energy convected by ions in the midplane. Finally, in order to fully substantiate this analysis, a quantitative analysis of transport is required in which the effect of shoulder formation on the perpendicular and parallel transport is described and compared with the evolution of the source and sink terms associated with the ionization front in the continuity and energy equations. These topics will be addressed in a dedicated follow-up paper.\\

\section{Conclusions}\label{Con}

Experiments carried out in AUG show how divertor collisionality, $\Lambda_{div}$, is the parameter triggering the shoulder formation in L-mode plasmas. A change in filamentary regime has also been measured for the first time in a fusion-relevant machine, linking the transition associated to filament disconnection from the wall to the formation of the density shoulder. These results lead to a scaling of $\lambda_n$ which is common for the three tokamaks of the “ITER Stepladder”: COMPASS, AUG and JET. An attempt has been carried out to extend these results into H-mode: first, the formation of a shoulder has been confirmed in inter-ELM H-mode plasmas. Second, the link between shoulder formation and filament transition has also been found to remain generally valid: collisionality remains the necessary condition for the shoulder formation, but a minimum level of deuterium fueling seems to be required on top of the collisionality threshold. Further experiments carried out in AUG show that the formation of the L-mode shoulder is also characterized by a strong cooling of the far SOL, which brings $T_i$ of both filaments and background to levels similar to those of $T_e$, $T_i \simeq T_e \simeq 20$ eV. This cooling is probably not the result of thermalization, although a more detailed analysis must be conducted before solid conclusions can be stated. A population of cold ions generated in the SOL mixing with the hot ions coming from the confined region represents a more likely explanation: for $\Lambda_{div}$ below unity, the distribution function of $T_i$ is centered around a single value, with a strong tail which can be associated to filaments. Instead, for $\Lambda_{div}>1$, the distribution is dominated by a peak of temperatures close to the Franck-Condon molecular dissociation energy of the $D_2$ molecules. This is consistent with the results of EMC3-EIRENE simulations, which predict a strong ionization front forming near the main chamber first wall. This is interpreted as the result of a drop in the neutral mean free path in the SOL, which, according to simulations, leads to a reduction of neutral density in the vicinity of the wall. This neutral layer would then act as a protective cushion for the wall, strongly reducing the average temperature of the ions reaching the wall (although not affecting the high energy tails of at least $T_i = 200$eV.  A common frame is proposed for all the presented data in which shoulder formation is described as the result of two feedback processes: first, the enhancement of filamentary transport associated with wall disconnection and regime transition, and triggered by $\Lambda_{div}$. Second, the opacity of the far-SOL, leading to a higher neutral density, increased ionization source, higher SOL density and even lower $\lambda_{mfp}$. The relation between the two processes depends critically on the importance of local ionization on the increase of filamentary transport for $\Lambda_{div} > 1$: while the formation of a shoulder without strong filamentary transport is considered unlikely, a shoulder could form in a SOL without high neutral densities if the increase of density associated with $S_{ion}$ is not critically required. This has implications for future devices: if a shoulder can be sustained without ionization, a two-region SOL would develop in which ions would first spread part of $P_{SOL}$ over relatively wide ranges and then be cooled down by the neutrals before arriving to the wall. If not, conditions in ITER and DEMO could be not favorable for the formation of a shoulder, despite the high $\Lambda_{div}$. In this case, the spread of energy would be similar to today's machines, and no cooling mechanism would prevent the hot ions from reaching the first wall. The proposed explanation of the shoulder formation provides a useful framework for the description of SOL transport, capable of including all presented data. This represents an important step in the progress towards a SOL model capable of quantitative predictions valid for ITER and DEMO.

\section*{Acknowledgements}

This work has been carried out within the framework of the EUROfusion Consortium and has received funding from the Euratom research and training programme 2014-2018 under grant agreement No 633053. The views and opinions expressed herein do not necessarily reflect those of the European Commission.

\end{document}